\documentclass{elsart}

\usepackage{amsmath,amssymb}

\begin{document}

\bibliographystyle{unsrt}

\begin{frontmatter}

\title{Continuous vs. discrete models for the quantum harmonic oscillator and the
hydrogen atom}

\author{Miguel Lorente\thanksref{email}}
\address{Departamento de F\'{\i}sica, Universidad de Oviedo, 33007 Oviedo, Spain}
\thanks[email]{E-mail: mlp@pinon.ccu.uniovi.es}
\begin{abstract} The Kravchuk and Meixner polynomials of discrete variable are
introduced for the discrete models of the harmonic oscillator and hydrogen atom.
Starting from Rodrigues formula we construct raising and lowering operators,
commutation and anticommutation relations. The physical properties of discrete models
are figured out through the equivalence with the continuous models obtained by limit
process.

\end{abstract}
\end{frontmatter}

\section{Introduction}

The method of finite difference is becoming more powerful in physics for different
reasons: difference equations are more suitable to computational physics and numerical
calculations; lattice gauge theories explore the physical properties of the
discrete models before the limit is taken; some modern theories have proposed physical
models with discrete space and time [1].

In recent papers [2], [3] we have presented the mathematical properties of
hypergeometric functions of continuous and discrete variable. We have worked out
general formulas for the differential/difference equation, recurrence relations,
raising and lowering operators, commutation and anticommutation relations. The
starting point is the general properties of classical orthogonal polynomials of
continuous and discrete variable [4] of hypergeometric type, in particular the
Rodrigues formula from which the raising and lowering operators are derived. Similar
results were obtained with more sofisticated method using the factorization of the
hamiltonian [6] [7]. In those papers the term ``oscillator" is used for the hamiltonian
that has equally spaced eigenvalues.

In this paper we study the physical properties of two simple examples, the harmonic
oscillator and the hydrogen atom on the lattice. We make the Ansatz that the discrete 
model has the same elements (hamiltonian,
eigenvalues, eigenvectors, expectation values, dispersion relations) as the continuous
one, provided the difference equation, the raising and lowering operator the
commutation and anticommutation relations become in the limit the equivalent elements
in the continuous case. To this scheme we can add the evolution of the fields with
discrete time, using some difference equation that replace Heisenberg equation. [8]

Some applications of discrete models have been presented elsewhere. Bijker et al. [9]
have apply the SO(3) algebra of the Wigner functions to theone-dimensional anharmonic
(Morse) oscillator. Bank and Ismail have apply Laguerre functions to the attractiv
Coulomb potential [10]

\section{The quantum harmonic oscillator of discrete variable}

We start from the orthogonal polynomials of a discrete variable, the Kravchuk
polynomials $K_n^{(p)} (x)$ and the corresponding normalized Kravchuk functions [1]
\begin{equation}
K_n^{(p)} (x) = d_n^{ - 1} \sqrt {\rho (x)\;} k_n^{(p)} (x),
\end{equation}
where $d_n^2  = \frac{{N!}}{{n!(N - n)!}}(pq)^n $ is a normalization constant, $\rho
(x) = \frac{{N!p^x q^{N - x} }}{{x!(N - x)!}}(pq)^n $ is the weight function, with $p >
0,\quad q > 0,\quad p + q = 1,\quad x = 0,1, \ldots N + 1.$

The Kravchuk functions satisfy the orthonormality condition
\begin{equation}
\sum\limits_{x = 0}^N {K_n^{(p)} (x)K_{n'}^{(p)} (x)}  = \delta
_{nn'},
\end{equation}
and the following difference and recurrence equations [2]:
\begin{multline}
 \sqrt {pq(N - x)(x + 1)} K_n^{(p)} (x + 1) \\ +  \sqrt {pq(N - x + 1)x} K_n^{(p)} (x
- 1)  + \left[ {x(p - q) - Np + n} \right] K_n^{(p)} (x) = 0,
\end{multline}
\begin{multline}
\sqrt {pq(N - n)(x + 1)} K_{n + 1}^{(p)} (x) \\ + \sqrt {pq(N - n + 1)n} K_{n -
1}^{(p)} (x) + \left[ {n(q - p) + Np - x} \right]K_n^{(p)} (x) = 0,
\end{multline}

From the properties of the Kravchuk polynomials we can construct raising and lowering
operators for the Kravchuk functions [2]
\begin{multline}
 L^ +  (x,n)K_n^{(p)} (x) = pq(x + n - N)K_n^{(p)} (x)  \\ + \sqrt {pq(N - x + 1)x}
K_n^{(p)} (x - 1) = \sqrt {pq(N - n)(n + 1)} K_{n + 1}^{(p)} (x),
\end{multline}
\begin{multline}
 L^ -  (x,n)K_n^{(p)} (x) = pq(x + n - N)K_n^{(p)} (x) \\ + \sqrt {pq(N - x)(x + 1)}
K_n^{(p)} (x + 1)   = \sqrt {pq(N - n + 1)n} K_{n - 1}^{(p)} (x).
\end{multline}
The raising operator satisfies
\[K_n^{(p)} (x) = \sqrt {\frac{{(pq)^n (N - n)!}}{{N!n!}}} \;\prod\limits_{k = 0}^{n -
1} {L^ +  (x,n - 1 - k)} K_0^{(p)} (x),
\]
where $K_0^{(p)} (x) = \sqrt {\frac{{N!p^x q^{N - x} }}{{x!(N - x)!}}} $
is the solution of the difference equation 
\[L^ -  (x,0)K_0^{(p)} (x) = 0.\]

It can be proved that the raising and lowering operators $L^ +  (x,n)$
 and $L^ -  (x,n)$ are mutually adjoint with respect to the scalar product (2). 

If we define the difference equation (3) as the operator equation $H(x,n)K_n^{(p)} (x)
= 0$ , we can factorize this equation as follows
\begin{align}
L^ +  (x,n - 1)L^ -  (x,n) &= pq(N - n + 1)n + pq(x + n - 1 - N)H(x,n), \\
L^ -  (x,n + 1)L^ +  (x,n) &= pq(N - n)(n + 1) + pq(x + n + 1 - N)H(x,n).
\end{align}

Now we make connection between the Kravchuk function and the Wigner functions that
appear in the generalized spherical functions [4]
\begin{equation}( - 1)^{m - m'} d_{mm'}^j (\beta ) = K_n^{(p)} (x),\;
\end{equation}
 where $j = {N \mathord{\left/
 {\vphantom {N 2}} \right.
 \kern-\nulldelimiterspace} 2}\;,\; m = j - n,\; m' = j - x,\; p = \sin ^2
\left( {{\beta  \mathord{\left/
 {\vphantom {\beta  2}} \right.
 \kern-\nulldelimiterspace} 2}} \right),\; q = \cos ^2 \left( {{\beta 
\mathord{\left/
 {\vphantom {\beta  2}} \right.
 \kern-\nulldelimiterspace} 2}} \right).$

Then formulas (3) to (8) can be written down in terms of the Wigner functions, namely,
\begin{eqnarray*}
\lefteqn{\frac{1}{2}\sin \beta \sqrt {(j + m')(j - m' + 1)} d_{m,m' - 1}^j (\beta )}\\
 &+&\frac{1}{2}\sin \beta \sqrt {(j - m')(j + m' + 1)} d_{m,m' + 1}^j (\beta ) +  
   (m - m'\cos \beta )d_{m,m'}^j (\beta ) = 0, 
\end{eqnarray*}\vspace{-3mm}\hfill (3a)
\vspace{-5mm}
\begin{eqnarray*}
\lefteqn{\frac{1}{2}\sin \beta \sqrt {(j + m)(j - m + 1)} d_{m - 1,m'}^j (\beta )} \\ 
 &+&  \frac{1}{2}\sin \beta \sqrt {(j - m)(j + m + 1)} d_{m + 1,m'}^j (\beta ) 
  - (m' - m\cos \beta )d_{m,m'}^j (\beta ) = 0,
\end{eqnarray*}\vspace{-3mm} \hfill (4a)
\vspace{-5mm}
\begin{eqnarray*}
L^ +  (m',m)\;d_{m,m'}^j (\beta) &=& \sin ^2 \frac{\beta }{2}(m +
m')\;d_{m,m'}^j (\beta ) \\
 & & + \frac{1}{2}\sin \beta \sqrt {(j - m')(j + m' + 1)}
\;d_{m,m' + 1}^j (\beta) \\
& =& \frac{1}{2}\sin \beta \sqrt {(j + m)(j - m + 1)} \;d_{m - 1,m'}^j (\beta ), 
 \end{eqnarray*}\vspace{-3mm}\hfill (5a)
\vspace{-5mm}
\begin{eqnarray*}
L^ -  (m',m)\;d_{m,m'}^j (\beta ) &=& \sin ^2 \frac{\beta }{2}(m +
m')\;d_{m,m'}^j (\beta ) \\
& &+\frac{1}{2}\sin \beta \sqrt {(j + m')(j - m' + 1)}
\;d_{m,m' - 1}^j (\beta ) \\ &=& \frac{1}{2}\sin \beta \sqrt {(j - m)(j + m + 1)}
\;d_{m + 1,m'}^j (\beta ). 
\end{eqnarray*}\vspace{-3mm}
	\hfill (6a)

 Notice that (3a) and (4a) are equivalent if we interchange $m \leftrightarrow m'$
 and take in account the general property of Wigner functions
\begin{equation}
d_{m,m'}^j (\beta ) = ( - 1)^{m - m'} d_{m',m}^j (\beta ).
\end{equation}
The same property is satisfied between (5a) and (6a). 

Expresions (5a) and (6a)
can be obtained directly from the properties of Wigner functions. In fact, it is known
[4, formula 5.1.19] that 
\begin{equation}
\frac{d}{{d\beta }}d_{m,m'}^j (\beta ) + \frac{{m' - m\cos \beta
}}{{\sin \beta }}d_{m,m'}^j (\beta ) = \sqrt {(j - m)(j + m + 1)} d_{m + 1,m'}^j
(\beta ),
\end{equation}
\begin{equation} - \frac{d}{{d\beta }}d_{m,m'}^j (\beta ) + \frac{{m' - m\cos \beta
}}{{\sin \beta }}d_{m,m'}^j (\beta ) = \sqrt {(j + m)(j - m + 1)} d_{m - 1,m'}^j
(\beta ),\end{equation}
The last equation (11), after interchanging $m \leftrightarrow
m'$ and using (9), can be transformed into 
\begin{equation*}
\frac{d}{{d\beta }}d_{m,m'}^j (\beta ) - \frac{{m - m'\cos \beta }}{{\sin \beta
}}d_{m,m'}^j (\beta ) = \sqrt {(j + m')(j - m' + 1)} d_{m,m' - 1}^j (\beta ).
\end{equation*}
\hfill (12a)

Combining (11) and (12a) we obtain (6a) and by similar method we obtain (5a) In
order to give a physical interpretation of the difference equation and raising and
lowering operators for the Kravchuk functions we take the limit when $N$
 goes to infinity and the discrete variable $x$
 becomes continuous $s$ . 

First of all, we take the limit of Kravchuk functions. We
write
\begin{eqnarray}
\nonumber\lefteqn{ K_n^{(p)} (x) = \left\{ {\frac{{n!(N - n)!}}{{N!(pq)^n
}}\frac{{(Npq)^n }}{{2^n (n!)^2 }}} \right\}^{{\raise0.7ex\hbox{$1$} \!\mathord{\left/
 {\vphantom {1 2}}\right.\kern-\nulldelimiterspace}
\!\lower0.7ex\hbox{$2$}}} \;\left\{ {\frac{{N!p^x q^{N - x} }}{{x!(N - x)!}}}
\right\}^{{\raise0.7ex\hbox{$1$} \!\mathord{\left/
 {\vphantom {1 2}}\right.\kern-\nulldelimiterspace}
\!\lower0.7ex\hbox{$2$}}} \;\left\{ {\left( {\frac{2}{{Npq}}}
\right)^{{\raise0.7ex\hbox{$n$} \!\mathord{\left/
 {\vphantom {n 2}}\right.\kern-\nulldelimiterspace}
\!\lower0.7ex\hbox{$2$}}} n!k_n^{(p)} (x)} \right\}} \\
& & = \xrightarrow[n \to \infty]{}
\left\{ {\frac{1}{{2^n n!}}} \right\}^{{\raise0.7ex\hbox{$1$} \!\mathord{\left/
 {\vphantom {1 2}}\right.\kern-\nulldelimiterspace}
\!\lower0.7ex\hbox{$2$}}} \;\left\{ {\frac{1}{{\sqrt {2\pi Npq} }}e^{ - s^2 } }
\right\}^{{\raise0.7ex\hbox{$1$} \!\mathord{\left/
 {\vphantom {1 2}}\right.\kern-\nulldelimiterspace}
\!\lower0.7ex\hbox{$2$}}} \;H_n (s) = \psi _n (s),
\end{eqnarray}
 where the last braket becomes the weight for the Hermite functions and the functions $\psi _n (s)$
 are the solution of the continuous harmonic oscillator (up to the constant
$(2Npq)^{{\raise0.7ex\hbox{${ - 1}$} \!\mathord{\left/
 {\vphantom {{ - 1} 4}}\right.\kern-\nulldelimiterspace}
\!\lower0.7ex\hbox{$4$}}} $ .

 In order to get the continuous limit of (4) we multiply it by ${2 \mathord{\left/
 {\vphantom {2 {\sqrt {2Npq} }}} \right.
 \kern-\nulldelimiterspace} {\sqrt {2Npq} }}$
 and substitute $x = Np + \sqrt {2Npq} \;s$ ; after simplification we get 
\begin{eqnarray*}
\lefteqn{\sqrt {\left( {1 - \frac{n}{N}} \right)2(n + 1)} \;K_{n + 1}^{(p)} (x) }Ê\\
& &\qquad + \sqrt {\left( {1 - \frac{{n - 1}}{N}} \right)2n} \;K_{n - 1}^{(p)} (x) -
2\left( {s +\frac{{(2p - 1)n}}{{\sqrt {2Npq} }}} \right)K_n^{(p)} (x) = 0.
\end{eqnarray*}

In the limit $N \to \infty $
 this equation becomes the familiar recurrence relation for the normalized Hermite
functions
\begin{equation}
\sqrt {2(n + 1)} \;\psi _{n + 1} (s) + \sqrt {2n} \;\psi _{n - 1} (s) = 2s\;\psi _n(s)
\end{equation}
 Before we take the limit of (5) we redifine the raising operator, substracting
from it one half equation (3), namely,
\begin{eqnarray}
\lefteqn{L^+(x,n)K_n^{(p)}(x)=\frac{1}{2}\Big\{ {\left[ {(x-Np)+n(p-q)} \right]}  K_n^{(p)}(x)} \nonumber\\
 & &\qquad  {-\sqrt {pq(N-x)(x+1)}K_n^{(p)}(x-1)+\sqrt {pq(N-x+1)x}K_n^{(p)}(x-1)} \Big\} \nonumber \\
 & &\qquad =\sqrt {pq(N-n)(n+1)}K_{n+1}^{(p)}(x)
\end{eqnarray}
We divide this expression by $h\sqrt {2Npq}\sqrt {Npq}$ where $h\sqrt {2Npq}=1$. After simplification we get

\begin{eqnarray}
\lefteqn{ \frac{1}{{\sqrt {Npq} }}\;L^ +  (x,n)K_n^{(p)} (x) =  \frac{1}{{\sqrt 2 }}\Bigg\{ \left[ {s + \frac{{n(p -
q)}}{{\sqrt {2Npq} }}} \right]K_n^{(p)} (x) } \nonumber\\
& &\qquad - \frac{1}{{2h}}\left[ {\sqrt {\left( {1 - \sqrt
{\frac{{2p}}{{Nq}}} s} \right)\left( {1 + \sqrt {\frac{{2q}}{{Np}}} s +
\frac{1}{{Np}}} \right)} K_n (x + 1)  } \right. \nonumber\\ 
& &\qquad  - {\left. {\sqrt {\left( {1 - \sqrt {\frac{{2p}}{{Nq}}} s + \frac{1}{{Nq}}}
\right)\left( {1 + \sqrt {\frac{{2q}}{{Np}}} s} \right)} K_n (x - 1)} \right]} \Bigg\} \nonumber\\
& &\qquad = \sqrt {\left( {1 - \frac{n}{N}} \right)(n + 1)} \;K_{n + 1}^{(p)} (x) \nonumber\\
& & \qquad =\xrightarrow [N \to \infty]{} \frac{1}{{\sqrt 2 }}\left\{ {s - \frac{1}{{2h}}\left[ {\psi _n (s + h) - \psi _n (s -
h)}
\right]} \right\} \nonumber\\
 & & \qquad =  \xrightarrow [h \to 0]{} \frac{1}{{\sqrt 2 }}\left\{ {s - \frac{d}{{ds}}} \right\}\psi _n (s) =
\;\sqrt {n + 1}
\;\psi _n (s),
\end{eqnarray}

Similarly from (6) we get
\begin{equation}
\frac{1}{{\sqrt {Npq} }}\;L^ -  (x,n)\;K_n^{(p)} (x)\xrightarrow[\stackrel{N \to \infty}{h \to 0}]{}\frac{1}{{\sqrt 2
}}\left\{ {s +
\frac{d}{{ds}}} \right\}\psi _n (s) = \;\sqrt n \;\psi _n (s).
\end{equation}
Therefore the raising and lowering operators for the Kravchuk functions become,
in the limit, creation and annihilation operators for the normalized Hermite
functions. 

We still have an other connection between the raising and lowering
operators of Wigner functions with the generators of the SO(3) algebra. 

From (5a) and (6a) we define
\begin{equation}
 A^ +  d_{mm'}^j (\beta ) \equiv \frac{1}{{\sqrt {Npq} }}\;L^ +  d_{mm'}^j (\beta ) =
\sqrt {\frac{{(j + m)(j - m + 1)}}{{2j}}} \;d_{m - 1,m'}^j (\beta ),
\end{equation}
\begin{equation}
 A^ - d_{mm'}^j (\beta ) \equiv \frac{1}{{\sqrt {Npq} }}\;L^ -  d_{mm'}^j (\beta ) =
\sqrt {\frac{{(j - m)(j + m + 1)}}{{2j}}} \;d_{m + 1,m'}^j (\beta ).
\end{equation}
Multiplying both expressions by the spherical harmonics $Y_{jm'}$ , adding for $m'$
 and using the property of Wigner functions 
\begin{equation}
Y_{jm}  = \sum\limits_{m'} {d _{m,m'}^j
\;Y_{jm'} }, 
\end{equation}
we obtain
\begin{equation}
A^ +  Y_{jm}  = \sqrt {\frac{{(j + m)(j - m + 1)}}{{2j}}} \;Y_{j,m - 1}  =
\frac{1}{{\sqrt {2j} }}J_ -  Y_{jm} 
\end{equation}
\begin{equation}
A^ -  Y_{jm}  = \sqrt {\frac{{(j - m)(j + m + 1)}}{{2j}}} \;Y_{j,m + 1}  =
\frac{1}{{\sqrt {2j} }}J_ + Y_{jm} 
\end{equation}
where $J_ +  ,J_ -$ are the generators of SO(3) algebra. 

For the commutation relations of these operators we have
\begin{equation}
(AA^ +   - A^ +  A)Y_{jm}  = \frac{m}{j}\;Y_{jm}  = \frac{1}{{2j}}2J_z Y_{jm}  =
\left( {1 - \frac{n}{j}} \right)Y_{jm} ,
\end{equation}

we substitute (20) in this expression and then take the limit:
\[
\left[ {AA^ +  } \right]d_{mm'}^j (\beta ) = \left( {1 - \frac{n}{j}} \right)d_{mm'}^j
(\beta )\xrightarrow[j \to \infty]{}\left[ {a,a^ +  } \right]\;\psi _n (s).
\]

For the anticommutation relation we have from (7) and (8):
\begin{equation}
\left( {AA^ +   + A^ +  A} \right)Y_{jm}  = \frac{1}{j}\left( {j(j + 1) - m^2 }
\right)Y_{jm}  = \frac{1}{j}\left( {\vec J^2  - J_z^2 } \right)Y_{jm}. 
\end{equation}
Again substituting (20) and taking the limit
\begin{eqnarray*}
\lefteqn{ \left( {AA^ +   + A^ +  A} \right)d_{mm'}^j (\beta )} \\
&& = \left\{ {(2n + 1) - \frac{{n^2
}}{j}} \right\}d_{mm'}^j (\beta ) \xrightarrow[j \to \infty]{}\left( {aa^ +   + a^ +  a} \right)\psi _n (s)   = (2n +
1)\psi _n (s)
 \end{eqnarray*}

This correspondence suggests that the operator algebra for the quantum harmonic
oscillator on the lattice is expanded by the generators of the SO(3) groups. The
commutation relations $\left[ {J_ +  ,J_ -  } \right] = 2J_z$ play the rol of the Heisenberg algebra, and the anticommutation relation
multiply by ${{\hbar \omega } \mathord{\left/
 {\vphantom {{\hbar \omega } 2}} \right.
 \kern-\nulldelimiterspace} 2}$ play the rol of the Hamiltonian. In order to complete the picture we define the
position and momemtum operators on the lattice as follows:
\begin{equation*}
 X:i\left( {\frac{\hbar }{{2M\omega }}} \right)^{{\raise0.7ex\hbox{$1$}
\!\mathord{\left/
 {\vphantom {1 2}}\right.\kern-\nulldelimiterspace}
\!\lower0.7ex\hbox{$2$}}} \left( {A + A^ +  } \right)d_{mm'}^j \left( \beta  \right) =
i\left( {\frac{\hbar }{{2M\omega }}} \right)^{{\raise0.7ex\hbox{$1$} \!\mathord{\left/
 {\vphantom {1 2}}\right.\kern-\nulldelimiterspace}
\!\lower0.7ex\hbox{$2$}}} \frac{{m' - m\cos \beta }}{{\sqrt j \sin \beta }}d_{mm'}^j
\left( \beta  \right) 
\end{equation*}
\vspace{-15mm}
\begin{eqnarray*}
\hspace*{-\mathindent} P:& &\left( {\frac{{M\hbar \omega }}{2}} \right)^{{\raise0.7ex\hbox{$1$}
\!\mathord{\left/
 {\vphantom {1 2}}\right.\kern-\nulldelimiterspace}
\!\lower0.7ex\hbox{$2$}}} \left( {A - A^ +  } \right)d_{mm'}^j \left( \beta  \right) = \left( {\frac{{M\hbar \omega }}{2}}
\right)^{{\raise0.7ex\hbox{$1$} \!\mathord{\left/
 {\vphantom {1 2}}\right.\kern-\nulldelimiterspace}
\!\lower0.7ex\hbox{$2$}}} \times \\
 \hspace*{-\mathindent} \times & & \left[ {\sqrt {\frac{{(j + m')(j - m' + 1)}}{{2j}}} d_{m,m' -
1}^j \left( \beta  \right) - } \right.\left. {\sqrt {\frac{{(j - m')(j + m' +
1)}}{{2j}}} d_{m,m' + 1}^j \left( \beta  \right)} \right] 
\end{eqnarray*}

with dispersion with respect to the state $ K_n^{(p)} (x)$
\begin{equation*}
\left( {\Delta X} \right)^2_n  = \left\langle {X^2 } \right\rangle _n  = \frac{\hbar
}{{2M\omega }}\left\langle {AA^ +   + A^ +  A} \right\rangle _n  = \frac{\hbar
}{{2M\omega }}\left( {2n + 1 - \frac{{n^2 }}{{j^2 }}} \right)
\end{equation*}
\vspace{-5mm}
\begin{equation*}
\left( {\Delta P} \right)^2_n  = \left\langle {P^2 } \right\rangle _n  = \frac{{M\hbar
\omega }}{2}\left\langle {AA^ +   + A^ +  A} \right\rangle _n  = \frac{{M\hbar \omega
}}{2}\left( {2n + 1 - \frac{{n^2 }}{{j^2 }}} \right)
\end{equation*}
from which the uncertainty relation follows:
\[
\left( {\Delta X} \right)_n\left( {\Delta P} \right)_n = \frac{\hbar }{2}\left( {2n + 1 -
\frac{{n^2 }}{{j^2 }}} \right)
\]
The eigenvalues of the Hamilton operator on the lattice are connected with the index
$m = j - n$ of the eigenvectors $d_{mm'}^j (\beta ).$ These eigenvalues are equally separated by
$\hbar \omega$ but finite $(m =  - j, \ldots  + j)$. The eigenvalues of the position operator
on the lattice are connected with the index
$m' = j - x$ of  $d_{mm'}^j (\beta )$. These eigenvalues are equally separated by $
\sqrt {\frac{\hbar }{{M\omega }}}$ but finite $(m' =  - j, \ldots  + j)$. Therefore the Planck
constant $\hbar$ plays a role with respect to the discrete space coordinate similar to the
discrete energy eigenvalues.

\section{Wave equation for the hidrogen atom with discrete variables} 

Our model is based on the properties of generalized Laguerre polynomials as continuous limit of the Meixner
polynomials of discrete variable. 

We start from the generalized Laguerre functions
\begin{equation}
\psi _n^\alpha  (s) = d_n^{ - 1} \sqrt {\rho _1 (s)} \;L_n^\alpha  (s)
\end{equation}
with $\alpha  >  - 1,\quad d_n^2  = \Gamma {{(n + \alpha  + 1)} \mathord{\left/
 {\vphantom {{(n + \alpha  + 1)} {n!,\quad \rho _1 (s) = s^{\alpha  + 1} e^{ - s}
,}}}
\right.
 \kern-\nulldelimiterspace} {n!,\quad \rho _1 (s) = s^{\alpha  + 1} e^{ - s} ,}}$
that satisfy the orthonormality condition
\begin{equation}
\int_0^\infty  {\psi _n^\alpha  (s)\;\psi _{n'}^\alpha  (s)\;s^{ - 1} ds = \rho
_{nn'} } 
\end{equation}
from the differencial equation and Rodrigues formula for the Laguerre polynomials [4]
we deduce the following properties for the Laguerre functions (25) 

i) differential equation
\begin{equation}
\psi _n^{\alpha ''} (s) + \left[ {\frac{\lambda }{s} - \frac{1}{4} - \frac{{\alpha ^2 
- 1}}{{s^2 }}} \right]\;\psi _n^\alpha  (s) = 0, \quad \lambda=n+\frac{1}{2}(\alpha+1).
\end{equation}
ii) Recurrence relations
\begin{eqnarray}
 \lefteqn{- \sqrt {\left( {n + \alpha  + 1} \right)(n + 1)} \psi _{n + 1}^\alpha  (s)} \nonumber \\
& & \qquad - \sqrt
{\left( {n + \alpha } \right)n} \psi _{n - 1}^\alpha  (s) + (2n + \alpha  + 1 -
s)\;\psi _n^\alpha  (s) = 0.
\end{eqnarray}
iii) Raising operator
\begin{eqnarray}
 L^ +  (s,n)\;\psi _n^\alpha  (s) &=&  - \frac{1}{2}(2n + \alpha  + 1 - s)\;\psi
_n^\alpha  (s) - s\frac{d}{{ds}}\;\psi _n^\alpha  (s) \nonumber \\ 
 & =&  - \sqrt {\left( {n + 1} \right)(n + \alpha  + 1)} \;\psi _{n + 1}^\alpha  (s) 
 \end{eqnarray}
iv) Lowering operator
\begin{eqnarray}
 L^ -  (s,n)\;\psi _n^\alpha  (s) &=&  - \frac{1}{2}(2n + \alpha  + 1 - s)\;\psi
_n^\alpha  (s) + s\frac{d}{{ds}}\;\psi _n^\alpha  (s) =  \nonumber\\ 
  &=&  - \sqrt {n(n + \alpha )} \;\psi _{n - 1}^\alpha  (s)
 \end{eqnarray}
from (29)and (30) we get the factorization of (27)
\begin{equation}
 L^ -  (s,n + 1)L^ +  (s,n) = (n + 1)(n + \alpha  + 1) - s^2 \left\{ {\frac{{d^2
}}{{ds^2 }} + \frac{\lambda }{s} - \frac{1}{4} - \frac{{\alpha ^2  - 1}}{{s^2 }}}
\right\}
\end{equation}
\begin{equation}L^ +  (s,n - 1)\;L^ -  (s,n) = n(n + \alpha ) - s^2 \left\{ {\frac{{d^2 }}{{ds^2 }}
+
\frac{\lambda }{s} - \frac{1}{4} - \frac{{\alpha ^2  - 1}}{{s^2 }}} \right\}
\end{equation}
Notice that (27) corresponds to some self adjoint operator of Sturm-Lioville type.
Also (29) and (30) are mutually adjoint operators with respect to the scalar product
(26). 

Similarly, in the discrete case, we defined the normalized Meixner functions
\begin{equation}
M_n^{(\gamma ,\mu )} (x) \equiv d_n^{ - 1} \sqrt {\rho _1 (x)} \;m_n^{(\gamma
,\mu )} (x)
\end{equation}
where $m_n^{(\gamma ,\mu )} (x)$
 are the Meixner polynomials, 
\[ d_n^2  = \frac{{n!\Gamma (n + \gamma )}}{{\mu ^n (1 - \mu )^\gamma  \Gamma (\gamma
)}},\quad \rho _1 (x) = \frac{{\mu ^x \Gamma (x + \gamma  + 1)}}{{\Gamma (x +
1)\Gamma (\gamma )}},
\]
and $\gamma ,\mu$ are some constants $0 < \mu  < 1,\quad \gamma  > 0$. The functions (33) satisfy the orthonormality condition
\begin{equation}
\sum\limits_{x = 0}^\infty  {M_n^{(\gamma ,\mu )} (x)\;} M_{n'}^{(\gamma ,\mu )}
(x)\frac{1}{{\mu (x + \gamma )}} = \rho _{nn'} 
\end{equation}
and the following properties: 

i) Difference equation
\begin{eqnarray}
\lefteqn{ \sqrt {\frac{{\mu (x + \gamma )(x + 1)(x + \gamma )}}{{x + \gamma  + 1}}} M_n (x + 1)}\nonumber  \\ 
&& \qquad + \sqrt {\mu (x + \gamma )x} M_n (x - 1) 
- \left[ {\mu (x + \gamma ) + x - n(1 - \mu )} \right]M_n (x)=0
\end{eqnarray}
ii) Recurrence relation
\begin{eqnarray}
- \sqrt {{\mu (n + \gamma )(n + 1)}} M_{n + 1} (x)  
 &-& \sqrt
{\mu (n + \gamma  - 1)n} M_{n - 1} (x) \nonumber \\
 &+& \left( {\mu x + \mu n + \mu \gamma  + n - x} \right)M_n (x) = 0 
\end{eqnarray}
iii) Raising operator
\begin{eqnarray}
 L^ +  (x,n)\;M_n (x) &=&  - \mu (x + \gamma  + n)\;M_n (x) + \sqrt {\mu (x + \gamma )x}
\;M_n (x - 1)  \nonumber\\ 
  &=& \sqrt {\mu (n + \gamma )(n + 1)} \;M_{n + 1} (x) 
\end{eqnarray}
 iii) Lowering operator
\begin{eqnarray}
 L^ -  (x,n)\;M_n (x) &=&  - \mu (x + \gamma  + n)\;M_n (x) \nonumber\\
 & & + \sqrt{\frac{{\mu (x + \gamma )(x + 1)(x + \gamma ) }}{{x + \gamma  + 1}}}\;M_n (x
+ 1) \nonumber\\ &=&  - \sqrt {\mu (n + \gamma  - 1)n} \;M_{n - 1} (x) 
\end{eqnarray}

A redefinition of (37) and (38) can be obtained substracting one half the difference equation
(35) from both:
\begin{eqnarray*}
\lefteqn{ L^ +  (x,n)M_n (x) =  - \frac{1}{2}\left( {2n\mu  + \mu \gamma  + n(1 - \mu ) +
(\mu  - 1)x} \right)M_n (x)}  \nonumber \\ 
  &&+ \frac{1}{2}\left\{ {\sqrt {\mu (x + \gamma )x} M_n (x - 1) - \sqrt {\frac{{\mu
(x +
\gamma )(x + 1)(x + \gamma )}}{{x + \gamma  + 1}}} M_n (x + 1)} \right\} \hfill (37a)
 \end{eqnarray*}
\vspace{-10mm}\begin{eqnarray*}
\lefteqn{ L^ -  (x,n)M_n (x) =  - \frac{1}{2}\left( {2n\mu  + \mu \gamma  + n(1 - \mu ) +
(\mu  - 1)x} \right)M_n (x) }\nonumber  \\ 
  &&+ \frac{1}{2}\left\{ {\sqrt {\frac{{\mu (x + \gamma )(x + 1)(x + \gamma )}}{{x +
\gamma  + 1}}} M_n (x + 1) - \sqrt {\mu (x + \gamma )x} M_n (x - 1)} \right\} \hfill (38a)
 \end{eqnarray*}

It can be proved that the difference equation (35)corresponds to some
self-adjoint operator of Sturm-Lioville type. Also the raising and lowering operators
(37) and (38) are mutually adjoint with respect to the scalar product (34).

The anticommutation relations for the raising and lowering operators (37a) and (38a)
are
\begin{eqnarray}
\lefteqn{\frac{1}{2}\left\{ {L^ +  (x,n - 1)L^ -  (x,n) + L^ -  (x,n + 1)\;L^ +  (x,n)}
\right\}} \nonumber \\
&=& \frac{1}{4}\left( {\mu \gamma  + (\mu  + 1)n + (\mu  - 1)x} \right)^2  \nonumber \\ 
&&  - \frac{\mu }{4}\Bigg\{ \sqrt {\frac{{\left( {x + \gamma } \right)^2 \left( {x + 1}
\right)\left( {x + \gamma  + 1} \right)\left( {x + 2} \right)}}{{x + \gamma  + 2}}}
\left( {E^ +  } \right)^2 - \left( {x + \gamma  - 1} \right) x \nonumber \\
&& - \left( {x + \gamma }
\right)\left( {x + 1} \right) + \sqrt {\left( {x + \gamma } \right)x\left( {x +
\gamma  - 1} \right)\left( {x - 1} \right)} \left( {E^ -  } \right)^2  \Bigg\} \nonumber  \\
 && + \frac{1}{2}(\mu +1)\Bigg\{ \sqrt {\frac{{\mu (x + \gamma )(x + 1)(x + \gamma )}}{{x +
\gamma  + 1}}} M_n (x + 1) \nonumber \\ 
&&- \sqrt {\mu (x + \gamma )x} M_n (x - 1) \Bigg\}
\end{eqnarray}

The commutation relations read

\begin{equation}
 L^ +  (x,n - 1)\;L^ -  (x,n) - L^ -  (x,n + 1)\;L^ +  (x,n) =  - \mu (2n + \gamma )
\end{equation}

In order to make connection between the Meixner functions of discrete variable (33) and
Laguerre functions of continuous variable (25) we substitute
$\gamma  = \alpha  + 1,\; \mu  = 1 - h,\; x = \frac{s}{h}$ and then take the limit $h \to
0$. 

The explicite expression of (33) in terms of $\alpha ,h$ , and $s$ reads
\begin{eqnarray*}
 \lefteqn{ M_n^{(\gamma ,\mu )} (x) = \sqrt {\frac{{\mu ^{n + 1} n!}}{{\Gamma (n + \gamma )}}}
\sqrt {\mu ^x \frac{{\Gamma (x + \gamma  + 1)}}{{\Gamma (x + 1)}}} \frac{{m_n^{(\gamma
,\mu )} (x)}}{{n!}} }  \\ 
 & & = \sqrt {\frac{{(1 - h)^{n + 1} n!}}{{\Gamma (n + \alpha  + 1)}}} \sqrt {\exp \left\{
{\frac{s}{n}\ln (1 - h)} \right\}h^{\alpha  + 1} \left( {x^{\alpha  + 1}  + O\left(
{\frac{1}{{x^2 }}} \right)} \right)} \frac{{M_n^{\alpha  + 1} \left( {\frac{s}{n}}
\right)}}{{n!}}
 \end{eqnarray*}
where we have used the asymptotic expansion
\begin{equation*}
\frac{{\Gamma (x + a)}}{{\Gamma (x + b)}} = x^{a - b} \left( {1 + O\left(
{\frac{1}{{x^2 }}} \right)} \right),\quad x \to \infty
\end{equation*}

In the limit we obtain (25), namely
\begin{equation*}
M_n^{(\gamma ,\mu )} (x)\xrightarrow [\substack{ h \to 0 \\ x \to \infty \\ hx \to s}]{}\sqrt {\frac{{n! }}{{\Gamma (n + \alpha 
+ 1)}}e^{ - s}s^{\alpha  + 1} } \;L_n^\alpha  (s) = \psi _n^\alpha  (s)
\end{equation*}
Also
\begin{eqnarray*}
(36) \quad   &\xrightarrow [h \to 0]{} &\quad (28) \\
(37a) \quad & \xrightarrow  [h \to 0]{} &\quad (29) \\
(38a) \quad& \xrightarrow [h \to 0]{} & \quad (30) \\
(39)\quad & \xrightarrow  [h \to 0]{} &\quad \frac{1}{2}\left [(30)+(31)\right ]
\end{eqnarray*}
In order to make application to the hydrogen atom we take the reduced radial equation
[5]
\begin{equation}
\frac{{d^2 u}}{{d\rho ^2 }} + \left[ {\frac{\nu }{\rho } - \frac{1}{4} -
\frac{{l(l + 1)}}{{\rho ^2 }}} \right]\;u(\rho ) = 0
\end{equation}
where
\[
\rho  \equiv \frac{{\sqrt {8M\left| E \right|} }}{\hbar }r\quad ,\quad \nu  =
\frac{{Ze^2 }}{\hbar }\sqrt {\frac{\mu }{{2\left| E \right|}}} 
\]

If we compair (41) with (27) we must have $
\rho  \equiv s,\;l(l + 1) = {{\left( {\alpha ^2  - 1} \right)} \mathord{\left/
 {\vphantom {{\left( {\alpha ^2  - 1} \right)} 4}} \right.
 \kern-\nulldelimiterspace} 4},$ hence, $\alpha  = 2l + 1,$ and $\nu  = n + \frac {(\alpha  +
1)}{2}= n + l + 1;$ since $l$ is integer, $\alpha $ and $\nu $ should be also integer. The energy
eigenvalues are given by
\begin{equation}
E_{\nu l}  =  - \frac{1}{2}\;\frac{{M\left( {Ze^2 } \right)^2 }}{{\hbar ^2
}}\;\frac{1}{{\nu ^2 }}\quad ,\quad \nu  = 1,2 \ldots 
\end{equation}

For fixed $\nu $ we still have degeneracy for $l = 0,1, \ldots \nu  - 1$

The corresponding eignvectors are given by (25), which after the substituion $\alpha  = 2l + 1,$
 and $n = \nu  - l - 1,$ become
\begin{equation}
\psi _{\nu l} (\rho ) = \left\{ {\frac{{(\nu  - l - 1)!}}{{(\nu  + l)!}}}
\right\}^{\frac{1}{2}} \rho ^{l + 1} \;e^{\frac{\rho}{2}} \;L_{\nu  - l - 1}^{2l + 1}\; (\rho)
\end{equation}
From the connection between the Meixner and Laguerre functions given above, we can make
the ansatz of a discrete model for the hydrogen atom where the reduced radial
equation is substituted by the difference equation (35) with $\gamma  = \alpha  + 1 = 2l + 2,\quad n = \nu  - l - 1,\quad x = 0,1,
\ldots $

The raising and lowering opertors (29), (30), respect to the radial quantum number $n$ , are substituted by the raising and lowering
operators (37) and (38) with respect to $n$ or $\nu  = n + l + 1.$

The anticommutation relations (32) + (31), which is proportional to hamiltonian of the
hydrogen atom (for the radial part), is substituted by the anticommutation realtions
(39). The commutation relations for the raising and lowering operators defining the Lie
algebra of the SU(1,1) group, are substituted by equation (40).

The expectation value of the discrete variable $x$ with respect to the Meixner functions $M_n^{(\gamma ,\mu )} (x)$
 is 
\begin{equation*}
\left\langle x \right\rangle _{n\gamma }  = \sum {M_n^{(\gamma ,\mu )} (x) \; x \;
M_n^{(\gamma ,\mu )} (x)},
\end{equation*}
that can be calculated with the help of the recurrence relation (36).

Finally the term $l(l + 1)$ in the Hamiltonian for the continuous and discrete case can be interpreted as the
eigenvalues of the Casimir operator for the SO(3) group, 
\begin{equation*}
\vec L^2  = \frac{1}{2}\left( {L^ +  L^ -   + L^ -  L^ +  } \right) + L_3^2 
\end{equation*}
where the operators $L^ +  ,\;L_ -$ y $L_3$ are acting on discrete space (see formulas (5) (6)).

\ack{
This work has been partially supported by D.G.I.C.Y.T. under contract PB-96-0538. The autor wants to express his gratitude to
Profs. A. Ronveaux and Y. Smirnov for valuable conversations.}

\end{document}